\documentclass[12pt]{iopart}

\def\um{\mu\mbox{m}}

\def\Wcm2{\mbox{W cm}^{-2}}
\def\Wcmum2{\mbox{Wcm}^{-2}\mu\mbox{m}^{2}}
\def\cm3{\mbox{cm}^{-3}}

\usepackage{graphicx}
\usepackage{color}
\usepackage{epsfig}
\usepackage{amssymb}

\begin{document}

\title{Radiation Pressure Acceleration by Ultraintense Laser Pulses}
\author{Tatiana V. Liseykina$^{1,2}$, Marco Borghesi$^{3}$, Andrea Macchi$^{3,4,5}$, Sara Tuveri$^{5}$}
\address{$^1$Max-Planck Institute for Nuclear Physics, Heidelberg, Germany}\ead{Tatyana.Liseykina@mpi-hd.mpg.de}
\address{$^2$Institute of Computational Technologies SD RAS, Novosibirsk, Russia}
\address{$^3$School of Mathematics and Physics, the Queen's University of Belfast, Belfast, UK}
\address{$^4$polyLab, CNR-INFM, Pisa, Italy}
\address{$^5$Dipartimento di Fisica ``Enrico Fermi'', Universit\`a di Pisa, Pisa, Italy}

\begin{abstract}
The future applications of the short-duration, multi-MeV ion
beams produced in
the interaction of high-intensity laser pulses with solid
targets will require improvements of
the conversion efficiency, peak ion energy, beam
monochromaticity and collimation.
Regimes based on Radiation
Pressure Acceleration (RPA) might be the dominant ones at ultrahigh
intensities and be most suitable for specific
applications. This regime may be reached already with present-day
intensities using circularly polarized (CP) pulses thanks to
the suppression of fast electron generation,
so that RPA dominates over sheath acceleration at any intensity.
We present a brief review of previous work on RPA
with CP pulses and a few recent results. Parametric studies in one
dimension were performed to identify the optimal thickness of
foil targets for RPA and to study the effect of a short-scalelength
preplasma. Three-dimensional simulations show the importance of
``flat-top'' radial intensity profiles to minimize the rarefaction of
thin targets and address the issue of angular momentum conservation
and absorption.
\end{abstract}

\pacs{52.38.Dx, 52.38.Kd, 52.38.Rr}

\submitto{\PPCF}

\maketitle

\section{Introduction}
In 1962, R.~L.~Forward considered the possibility of interstellar travel by 
a rocket propelled by an Earth-based laser beam 
(see \cite{forwardJS84} and references therein).
The concept is as simple as follows: the rocket's engine and fuel are replaced 
by a light sail, i.e. a mirror, and the force exerted on the sail due to 
the radiation pressure of the laser light boosts the rocket.
In 1966, G.~Marx \cite{marxN66} found that the \emph{efficiency} of the system,
i.e. the ratio between the mechanical energy of the object accelerated by
the laser beam and the energy contained in the laser beam itself, would
approach unity as the velocity of the object approaches the speed
of light.
An heuristic (though incomplete) argument might be given in terms of light
quanta, i.e. photons, although the system can be described as entirely
classical: let us consider a ``perfect'' mirror irradiated by a
monochromatic light wave of frequency $\omega$,  that contains a
certain number $N$ of
photons and thus has a total energy $N\hbar\omega$. The mirror reflects
photons conserving their number in any reference frame. If the mirror has an
(instantaneous) velocity $V=\beta c$ in the laboratory frame, the
frequency of the reflected photons is $\omega'=\omega(1-\beta)/(1+\beta)$.
Thus, since $N$ is invariant, the energy of the reflected pulse tends to
zero if $V\rightarrow c$, so that a mirror moving at a speed close to $c$
absorbs almost all the energy of the incident pulse.

Marx paper's conclusions turned out to be right although its approach needed
a critical revision, as can be found in the rigorous and
pedagogical description of Ref.\cite{simmonsAJP92}.
According to the formulas in Ref.\cite{simmonsAJP92}.
it would take a three years for a 10~TW laser to accelerate a
$10^3~\mbox{kg}$ rocket to $V=(4/5)c$. Scaling this result
to the typical parameters of superintense laser pulses and micro-targets,
we obtain that about $5\times 10^{10}$ Carbon ions might be accelerated
to the same speed in 1~ps by a 1~PW laser, which is within the capabilities
of present technology. 
Notice that the required acceleration length would
be of the order of $100~\um$, which is a suitable value for the Rayleigh 
length, so that laser pulse diffraction should not be a strong limiting
factor on the achievable energies. This makes the perspective of
Radiation Pressure Acceleration (RPA) attractive for applications requiring
large numbers of relativistic ions.

Nearly all of the experiments reported in the last decade on the acceleration
of ions (mainly protons) by superintense laser pulses (see e.g.
\cite{borghesiFST06,mckennaPPCF07} and refs. therein)
are not based on RPA
but instead on the Target Normal Sheath Acceleration (TNSA) mechanism,
in which ions are accelerated by space-charge fields created by
multi-MeV electrons escaping into vacuum.
The dominance of RPA over TNSA in thin solid targets irradiated
at intensities higher than those of present-day experiments
has been claimed by Esirkepov and coworkers
\cite{esirkepovPRL04,esirkepovPRL06},
on the basis of simulations showing a transition occurring at some intensity
value above $10^{21}~\Wcm2$, with a strong dominance leading to the
so-called ``piston'' regime over $10^{23}~\Wcm2$. Such an intensity may be
available only in several years from now thanks to the development of
advanced laser facilities. Experimentally, a preliminary indication of RPA
effects in thin targets at intensities approaching $10^{20}~\Wcm2$
has been published recently \cite{karPRL08} (some experimental results on
ion acceleration at intensities $\leq 10^{18}~\Wcm2$
were also interpreted in terms of purely ponderomotive, i.e. radiation pressure
effects \cite{badziakPPCF04}).

The question then arises whether it is possible
to achieve a RPA-dominated regime already at lower intensity; this corresponds
in practice to quench the generation of high-energy electrons which
drive TNSA but do not contribute to RPA. This may be possible
using circularly polarized (CP) laser pulses at normal incidence 
because the \emph{oscillating}
components of the Lorentz force in the direction perpendicular to
the sharp density gradient vanish (for a plane wave) or are relatively small
(for a finite laser spot size): as a consequence, the motion of electrons
at the interaction surface is predominantly adiabatic and electron heating 
is strongly reduced, while the space-charge field created to balance the 
local radiation
pressure (i.e. the ponderomotive force) accelerates ions.

The strong differences between the cases of linearly polarized (LP) and CP
pulses have been evidenced in some
papers by our group \cite{macchiPRL05,liseikinaAPL07}, mostly for the
case of ``thick'' targets, i.e. thicker than the skin layer. 
These studies showed that for CP the interaction
accelerates all the ions in the skin layer and the fastest ones produce a
very dense ``bunch'' with a narrow energy spectrum, directed in the
forward direction.

Recently, the experimental availability of ultrathin targets (i.e. a with
thickness down to a few nanometers) and high-contrast laser pulses
(see e.g. \cite{neelyAPL06,thauryNP07,ceccottiPRL07})
has calles for studies of CP-RPA with such pulses and targets. The simulations
performed independently by several groups
\cite{zhangPP07,zhangPP07b,robinsonNJP08,klimoPRSTAB08,yanPRL08}
suggest that indeed
the whole target may be accelerated, leading to efficient generation of
large numbers of ions with monoenergetic spectra in the near-GeV range.
Presently, no experiment using CP pulses at normal incidence has been
reported in publications yet, but several related proposals have been made,
so that the CP-RPA concept is expected to be explored soon.

The present paper reviews the main issues of CP-RPA and reports novel
numerical results on parametric studies in one spatial dimension (1D),
showing the role of the target thickness and the case of RPA in
short-scalelength preformed plasmas, as well as first results in fully 3D
geometry where, in particular,
the issue of angular momentum conservation can be addressed.

\section{Theory and earlier work}
Firstly we shall provide a brief description of the interaction in the case 
of a thick target. The ponderomotive force of the laser pulse causes 
the electrons to pile up in the skin layer until it is balanced by the
charge separation field that accelerates the ions.
The ions produce a sharp density spike at the end of the
skin layer where hydrodynamical breaking occurs, with the
faster ions creating a dense bunch (with a narrow spectrum) that
moves ballistically into the plasma
(a rather similar dynamics has been noted in the case of radial ponderomotive
acceleration of ions in an underdense plasma \cite{macchiPPCF08}). 
A detailed physical description and a simple model of RPA in thick targets
(assuming a priori non--relativistic ions)
have been given in previous work \cite{macchiPRL05,liseikinaIEEE08}.
The model provides the following scaling, valid for sub-relativistic 
ion velocities, for the maximum ion velocity 
(which almost corresponds to the bunch velocity)
and the corresponding energy:
\begin{equation}
\frac{v_{im}}{c} = 2\sqrt{\frac{Z}{A}\frac{m_e}{m_p}\frac{n_c}{n_e}}a_L,
\qquad
{\cal E}_m=\frac{m_i}{2}v_{im}^2=2m_e c^2Z\frac{n_c}{n_e}a_L^2,
\label{eq:scaling}
\end{equation}
where $n_c=m_e\omega^2/4\pi e^2=1.1 \times 10^{21}~\mbox{cm}^{-3}$
is the cut--off density for the laser wavelength $\lambda=2\pi c/\omega$,
$n_e$ is the background electron density,
$a_L=eE_L/m_e\omega c=0.85 (I\lambda^2/10^{18}~\mbox{W cm}^{-2})$
is the dimensionless amplitude of the laser pulse
with electric field $E_L$ and intensity $I$, and other symbols are standard.
Actually, this result has been derived in the limit of relatively low
intensity, bus this scaling has been found to hold up
to much higher intensity in parametric 1D simulations \cite{liseikinaIEEE08}.

The ion bunch is formed in a time of the order of $\sim c/\omega_p v_{im}$
(where $\omega_p$ is the plasma frequency)
after which it exits the skin layer accompanied by
neutralizing electrons, and the laser pulse may accelerate a new layer.
If energies higher than the above estimate must be reached, it is necessary to
repeat the acceleration stage on the same ions, i.e. the target must be thin
enough in order to bunch and accelerate all ions via several
cycles. 
Simulation results on the acceleration of ultrathin targets have been
interpreted with the model of the accelerating mirror \cite{robinsonNJP08}
(where the mirror is assumed to be a ``rigid'' object, neglecting any
internal dynamics).

Although a few authors have proposed the RPA of a thin foil
as a way to generate high-energy protons, this approach seems to be most
interesting for the acceleration of higher-$Z$ ions. In fact, while it seems
technologically unfeasible to have an ultrathin target made of hydrogen
only, in a target made of multiple species all the ions will be accelerated 
to the same velocity, resulting in higher energies for the heavier species.
If a lighter species (e.g., hydrogen) is present, these ions will be first
accelerated overtaking the heavier ones. This will cause them to decouple
from the laser pulse, which is screened by the heavier ion layer.
The heavier ions will be accelerated until they reach the lighter ions,
allowing the laser to reach them and accelerate them further. In the end
all species will have the same velocity.
Due to this effect, for an ultrathin target of a single 
material (e.g. Carbon) a monoenergetic ion spectrum is expected.

The above picture
of the acceleration dynamics should change when the
ions finally reach a speed close to $c$, as they will no longer be separated
from the electrons. 
The ``laser-piston'' regime investigated by Esirkepov et al.
\cite{esirkepovPRL04} corresponds to conditions in which the
ions are promptly accelerated to relativistic velocities and stick to the
electrons, which may not be assumed to be in a mechanical quasi-equilibrium
anymore. In the present paper we restrict our analysis to the regime of
non-relativistic ions because near-term experiments on RPA are unlikely to
have the potential to accelerate ions up to strongly relativistic energies.

The theoretical picture and the predicted scalings have been supported
by 1D simulations. So far multi-dimensional effects have been addressed
at most by 2D simulations for both thick
\cite{macchiPRL05,liseikinaAPL07,liseikinaIEEE08}
and thin \cite{robinsonNJP08,klimoPRSTAB08,yanPRL08} targets.
In the thick target cases, the energy spectrum is basically determined by
the convolution of the 1D scaling law with the intensity profile of the pulse.
The angle of emission of ions is energy-dependent, but a good
collimation is already obtained for a Gaussian pulse profile
\cite{liseikinaAPL07}. For thin targets, the use of 
a flat-top profile increases monoenergeticity and collimation
and quenches the heating of electrons,
as expected \cite{robinsonNJP08,klimoPRSTAB08}.

The issue of target stability during RPA has been addressed in thin foil
2D simulations for CP \cite{klimoPRSTAB08} and also for linear polarization
in the ultraintense regime \cite{pegoraroPRL07}, showing a
bending instability which has been interpreted to be of the Rayleigh-Taylor
type and hence can be described in terms of the radiation pressure only.
Simulations for thick targets, however, have shown that surface instabilities
are weaker for CP pulses than for linearly polarized ones. The quality of
the ion beam is expected to be lower for linear polarization
\cite{liseikinaIEEE08}.

\section{1D simulations}

\subsection{The role of the target thickness}

\begin{figure}
\begin{center}
\includegraphics[width=0.55\textwidth]{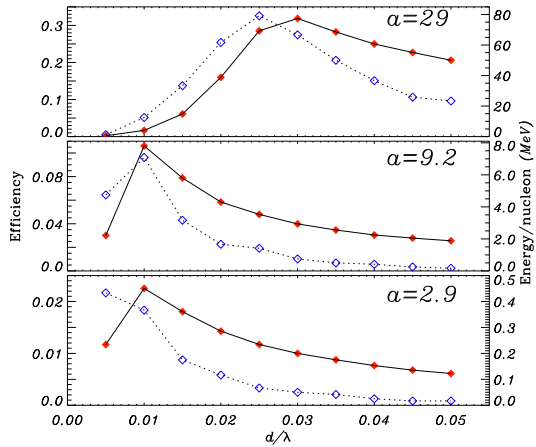}
\caption{Energy conversion efficiency into ions (red
filled diamonds) and the ``peak'' energy per nucleon (blue empty
diamonds) as a function of the target thickness,
investigated by parametric 1D PIC simulations. The top, middle and bottom
plots are for a pulse amplitude $a=29$, $9.2$ and $2.9$, respectively.
In all the runs, the laser pulse had a duration of $9$ cycles (FWHM), the
electrons density was $n_e=250n_c$ and the charge-to-mass ratio was
$Z/A=1/2$.
\label{fig:param_thick}}
\end{center}
\end{figure}

We performed a parametric study to determine the optimal values
of target thickness $d$
to obtain higher efficiency and/or ion energy for given laser parameters.
In order to be able to cover a quite wide range of parameters and
to simulate ``realistic'' target densities we used 1D simulations, which
in the CP-RPA regime have so far proved to yield efficiency and ion
energy values close to those from 2D or 3D simulations for those cases
where a comparison is possible (i.e. for moderate density values).
Results are shown in Fig.~\ref{fig:param_thick}. The electron density
of the target and the pulse duration were kept constant for all runs
and corresponded, for a laser wavelength $\lambda=0.8~\um$, to
$n_e=4.3 \times 10^{23}~\cm3$ and $\tau_L=24~\mbox{fs}$.
The three values of the dimensionless amplitude that were studied
($a=$2.9, 9.2 and 29) corresponded to intensities
$I=1.8 \times 10^{19}$, $1.8 \times 10^{20}$ and $1.8 \times 10^{21}~\Wcm2$,
respectively.

The values of $d$ for which efficiency and ion energy have their maximum are
close to each other and, as expected, they correspond to ultrathin,
sub-micrometer targets. The strong decrease of efficiency and energy
for smaller values of $d$ may be explained with the onset of relativistically
induced transparency in the thin foil when $d\simeq\lambda a(n_c/n_e)$
\cite{vshivkovPP98}, so that the total radiation pressure on the target
decreases. This point will be further discussed below when 
three-dimensional effects are addressed (section \ref{sec:3D}).

The energy per nucleon reported in Fig.~\ref{fig:param_thick}
can be scaled to all species with
$Z/A=1/2$. For Carbon ($A=12$) the highest energy of $0.96~\mbox{GeV}$
is obtained for $a=29$ and $d=0.025\lambda$.
Notice that these are ``peak'' energies which correspond to a distinct
maximum in the ion spectra. However, depending on the interaction parameters
some tail of higher energy ions appears. Moreover, the width of the ion
energy peak also varies troughout the simulations and does not remain constant
in time, as some broadening is observed after the laser pulse is
over. This broadening appears to be related with electron heating which
occurs at the end of the acceleration stage, creating ``warm'' electrons
which are much less energetic than those produced for LP interaction but
may already drive the expansion of the thin plasma foil. Hence,
monoenergeticity of ions appears to be a non-trivial issue already in 1D.

In higher dimensionality it is known that the intensity distribution in
the laser spot gives rise to an energy spread correlated with the
direction of laser-accelerated ions \cite{liseikinaAPL07}, so that a
``flat-top'' distribution, whenever feasible, may improve
monoenergeticity as well as beam collimation (see e.g. 2D simulations
in Ref.\cite{robinsonNJP08}). Additional effects of the pulse profile are
also discussed in section \ref{sec:3D}.

\subsection{RPA in preformed plasmas}

The use of ultrathin targets in experiments will require the use of systems
with an extremely high contrast ratio, otherwise the prepulse preceding
the main interaction pulse will destroy the target completely.
Interaction experiments in such a regime appears to be presently possible
\cite{neelyAPL06,ceccottiPRL07},
thanks e.g. to the use of plasma mirrors to improve the contrast
\cite{thauryNP07}.
Such conditions are optimal to test the CP-RPA of
ultrathin targets, provided that the strategies implemented to improve
the pulse contrast are compatible with preserving the circular polarization
of the pulse. It is also worth to stress that the need of normal pulse
incidence might also be non-trivial to be experimentally satisfied
due to the danger of back-reflection from the overdense plasma.

It is interesting in any case to consider the possibility of the interaction
of the CP pulse with a non-uniform preplasma, as this may be present
in experiments where the contrast ratio is modest.
Moreover, the expected scaling of the ion energy with the inverse of the
plasma density suggests that, in a preformed plasma, a given laser pulse
may produce a lower total number of ions but with higher energy, as the
interaction occurs with the layer at the cut-off density $n_c$ which is
typically less than one hundredth of the solid density.

\begin{figure}
\includegraphics[width=0.49\textwidth]{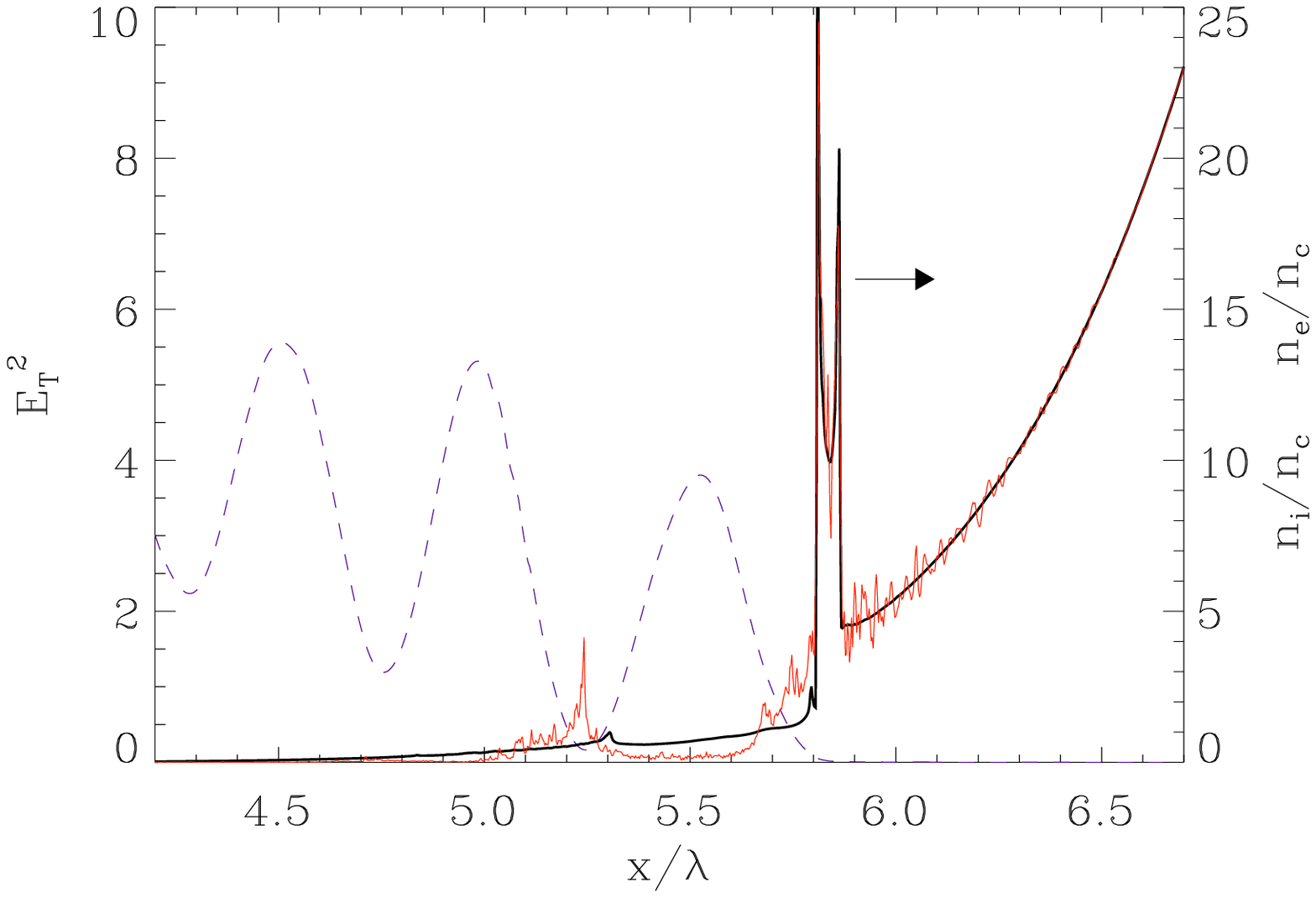}
\includegraphics[width=0.49\textwidth]{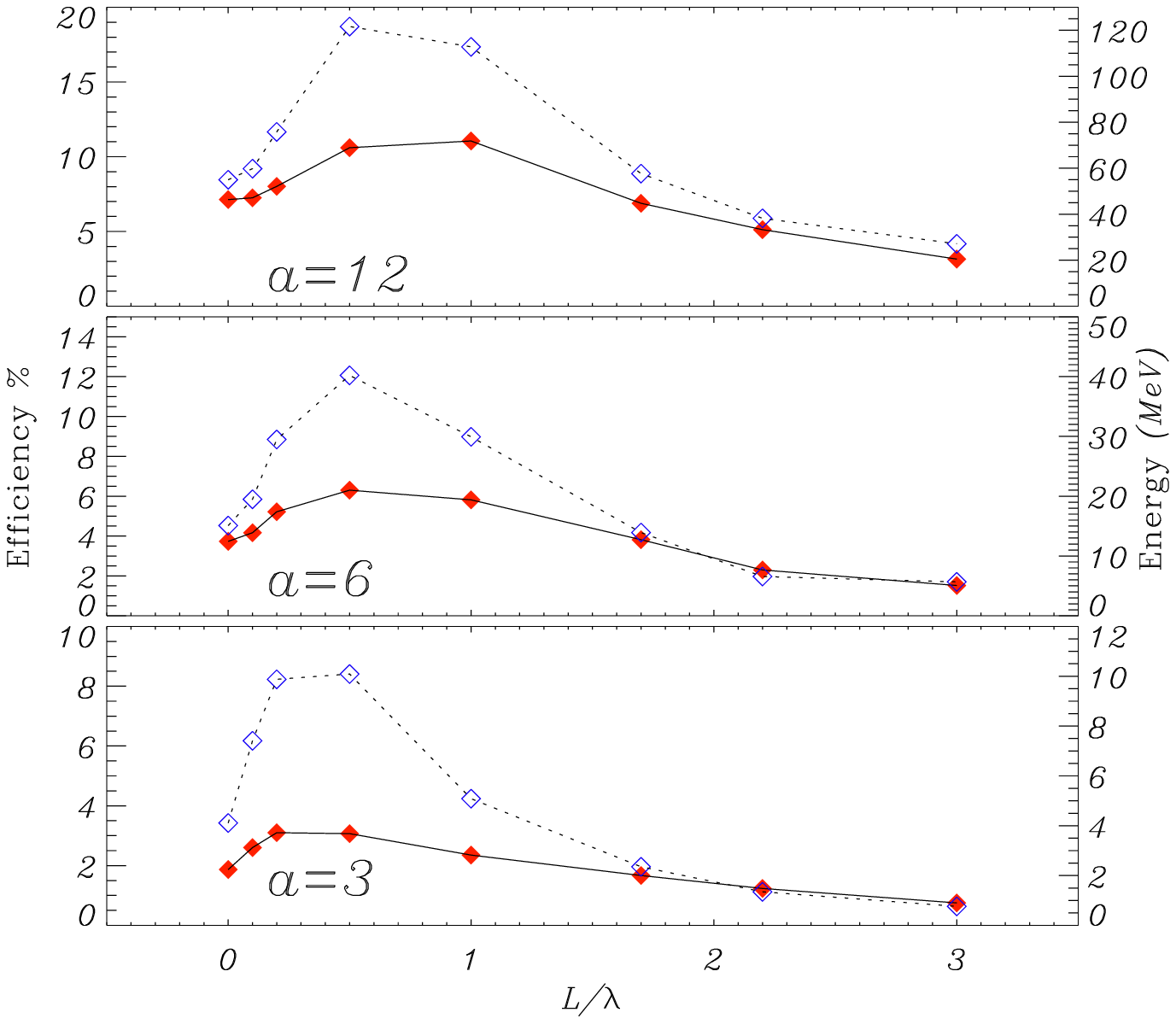}
\caption{Interaction with preformed plasmas. Left: snapshots of the profiles
of $E^2_T=E^2_y+E^2_z$ (dashed blue line), $n_i$ (thick black line) and
$n_e$ (thin red line) soon after the formation of the ``fast'' ion bunch
(evidenced by the arrow). The pulse intensity was $a=3$
corresponding to  $1.2\times 10^{19}~\mbox{W cm}^{-2}$
for $\lambda=1~\mu\mbox{m},$ the density profile was rising with a $\sim (x-x_0)^9$ law up to
a peak density $n_0=40n_c.$
Right: conversion efficiency (red filled diamonds, solid line)
and peak energy (blue empty diamonds, dashed line) of ions
as a function of the density scalelength $L$, for three values of the
laser amplitude $a$.
In all the simulations the laser pulse had a duration of $9$ cycles (FWHM)
and the density profile was rising with a $\sim (x-x_0)^4$ law up to
a peak density $n_0=16n_c$, and then remained constant.
\label{fig:param_preplasma}}
\end{figure}

We performed a set of parametric 1D simulations assuming initial density
profiles of power-law type (i.e. $n_0(x) \sim (x-x_0)^k$ for $x>x_0$)
and different values of the density scalelength at the cut-off layer,
$L=n_c/|\partial_x n_0|_{n_0=n_c}$. The snapshot of the ion density $n_i$ 
in Fig.\ref{fig:param_preplasma} shows that the $n_i$ undergoes spiking 
and ``breaking'' and that a ``fast'' bunch forms
near the cut-off density layer,
with features very similar to the case of a sharply rising density (no
preplasma) \cite{macchiPRL05}. The bunch density is several times $n_c$.
As a function of $L$, both the maximum ion energy and the conversion
efficiency have their maxima for a very short scalelength
$L \simeq 0.5\lambda$,
as also shown in Fig.\ref{fig:param_preplasma} where $L=0$ corresponds
to the case of no preplasma, i.e. a step-like profile.
When compared to the energy scaling (\ref{eq:scaling}), the observed
ion energy would correspond to a density value intermediate between $n_c$
and the peak density ($16n_c$) of the
profile. The decreasing of energy and efficiency for larger values of $L$ might
be related to the weaker
coupling of the laser pulse to the cut-off layer;
a relevant part of the pulse energy is found to be absorbed in the
underdense plasma, e.g. by excitation of plasma waves, causing the
absorption degree into electrons to be higher than into ions, and 
consequently decreasing the total radiation pressure.
The stronger heating of electrons may account for the
broad energy spectrum that is observed for non-optimal values of $L$; 
Fig.\ref{fig:param_preplasma}
reports the maximum or cut-off energy, but several and broad ``peaks''
may appear in the spectrum under such conditions, while the spectrum is
narrow for the case of absolute maximum energy.
These preliminary results suggest that RPA may be strongly affected
by prepulse effects. However this may allow 
to achieve dense bunches of multi--MeV ions
using ultrashort pulses with controlled contrast. For longer pulses (hundreds
of fs), this approach may become ineffective because of strong steepening
effects during the rise of the pulse, decreasing the value of $L_c$.
The width of the target layer that remains undamaged by the prepulse may
also play a role because the ions may undergo relevant collisional losses 
while crossing the solid-density region (see e.g. the discussion in
\cite{karPRL08}).

\section{3D simulations}
\label{sec:3D}

As it is always the case for computational plasma physics, 3D simulations
would be required for a ``realistic'' description, but the limits of
computing power forces the restriction to a narrow set of ``feasible''
parameters. This is the case for CP-RPA where, furthermore, the resolution
must be high enough to resolve effects such as the strong spiking of
the density observed in 1D and 2D.
Thus, only a few 3D runs could be performed and for plasma densities much less
than solid-density values, though well above $n_c$.

The comparison with 1D and 2D results is important also
because a CP pulse carries a net angular momentum whose
conservation law appears as an additional constraint in 3D.
Note that, despite the strong absorption of pulse energy, no absorption 
of angular momentum is expected, at least as long as the acceleration is 
adiabatic (as it is assumed in the ```perfect mirror" model).
In fact, coming back to the heuristic argument of the introduction, if the
number of photons is conserved and the reflected beam conserves
helicity (which  can be shown to hold), no angular momentum is
left in the target because the ``spin''  of any photon is $\hbar$, independent
on the frequency.

\begin{figure*}
\includegraphics[width=0.31\textwidth]{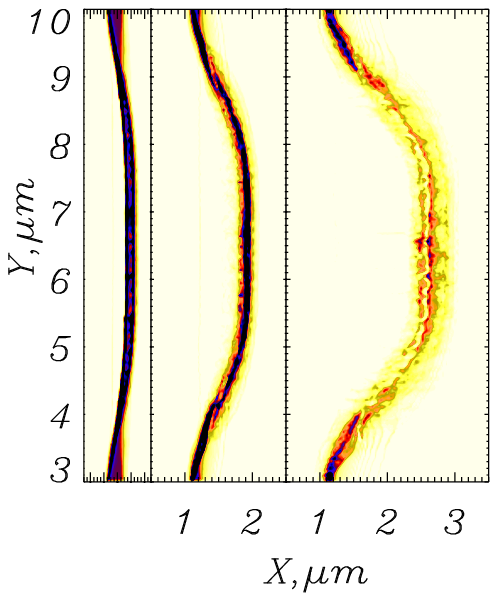}
\includegraphics[width=0.25\textwidth]{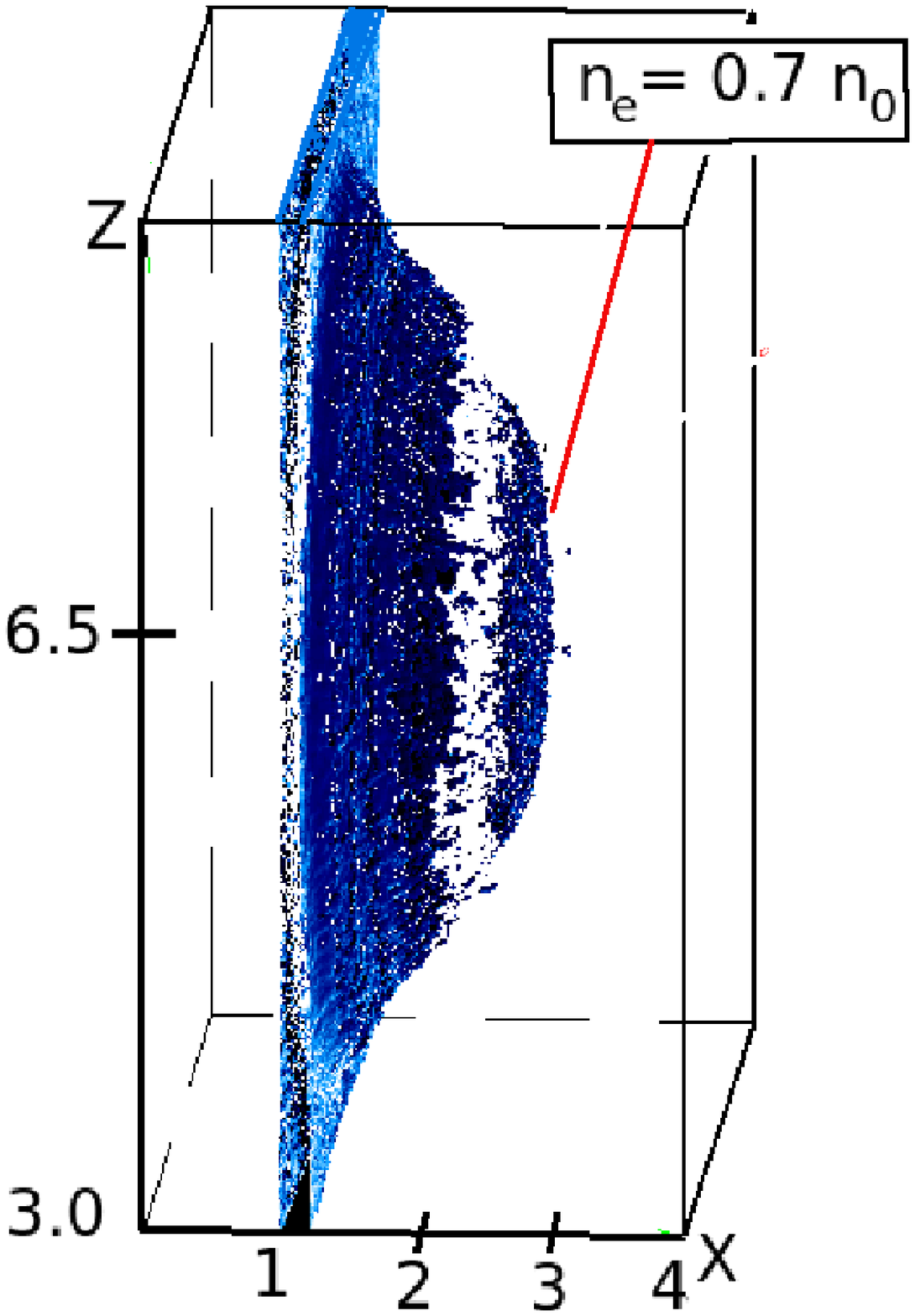}
\includegraphics[width=0.4\textwidth]{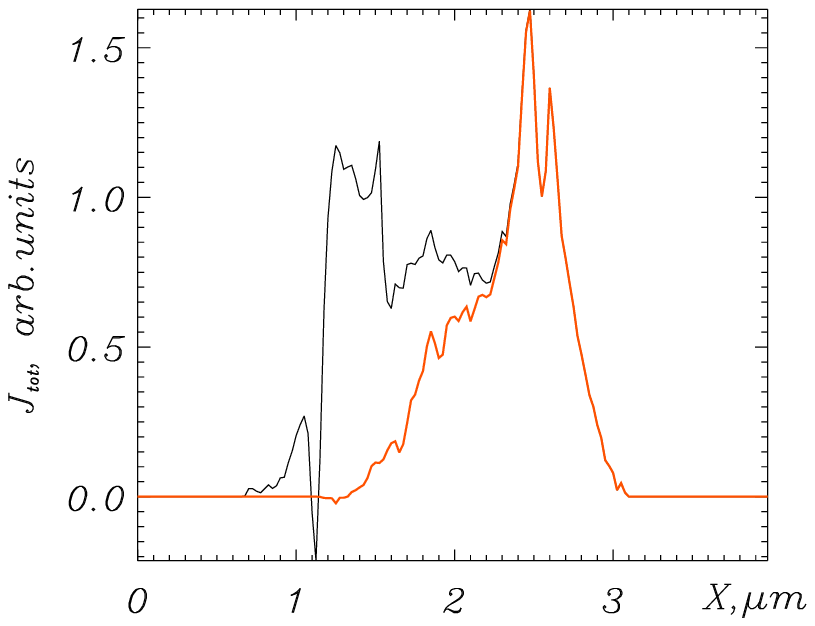}
\caption{(a) The distribution of ion density at $t=75, 100$ and $130$ fs in $(xy)$ plane, (b) the distribution of the ion density at $t=130$ fs (the laser pulse is over), (c) the integral of the ion poloidal current $\oint J_{\varphi} dy dz$ at $t=130$ fs versus longitudinal coordinate, showing the angular momentum absorption: black linefor the region with radius $r=4.5~\mu$m around the $x$-axis;
red line for the  $r=2.5~\mu$m region around the $x$-axis.
\label{fig:3D1}}
\end{figure*}

In order to study these issues in details, we performed several 3D simulations.
In all the runs discussed here the normally incident laser pulse was
circularly polarized with a peak intensity of 
3.4$\times$ 10$^{19}$~W cm$^{-2}$ and $\sim$ 60~fs duration and the target
consisted of electrons and protons.
Figure~\ref{fig:3D1} presents the results of the interaction of
a laser pulse
with a ``flat-top'' intensity profile of 6 $\mu$m width
with a target
of density $n_e=16 n_{cr}=1.7\times 10^{22}$ cm$^{-3}$ and 0.3 $\mu$m thick.
Figure~\ref{fig:3D1}~(a) shows the projected 2D distributions of ion density 
at $t=75, 100$ and $130$ fs, and Fig.~\ref{fig:3D1}~(b) --
the 3D plot of the ion density when the
laser pulse is over. The density of the ``bunch'' is approximately 0.7 of the
initial density of the target, the peak energy of ions in this bunch is 
$\sim 4$ MeV, the number of accelerated ions is $\sim 4\times10^{10}.$
When the laser pulse is over most of the absorbed angular
momentum ($\sim 4\%$) is transferred to the ions 
(the energy absorption in this case was $\sim 7\%$).
To prove that a torque on the plasma ions exists,
we plot in Fig.~\ref{fig:3D1}~(c) the integral (over $(y,z)$) of the 
poloidal current $J_{\varphi}$ of the ions.
We thus see that on the average there is a net ``rotation'' of the
ions, while the same plot for the electron current shows that the latter
averages over $x$ almost to zero.
The poloidal ion current is concentrated near the edge of the laser spot
where the angular momentum density has a maximum.

Angular momentum absorption in laser-plasma interactions has been mostly
studied so far in underdense plasmas as a problem closely related to
the generation of a steady magnetic field (Inverse Faraday Effect).
Haines \cite{hainesPRL01} reported a short critical review of previous work
and discusses effects leading to a torque on the plasma ions.
The issue of angular momentum absorption in overdense plasmas has received
much less attention so far. In the present context, 
the observation of some degree of angular momentum
absorption is a signature of non-adiabatic or ``dissipative'' effects 
(which are an interesting issue in collisionless systems)
not included in the ``perfect mirror'' model of RPA.
They may be related to the onset of hydrodynamical breaking during the
acceleration process \cite{macchiPRL05}, violating the adiabaticity condition.

\begin{figure*}
\includegraphics[width=0.45\textwidth]{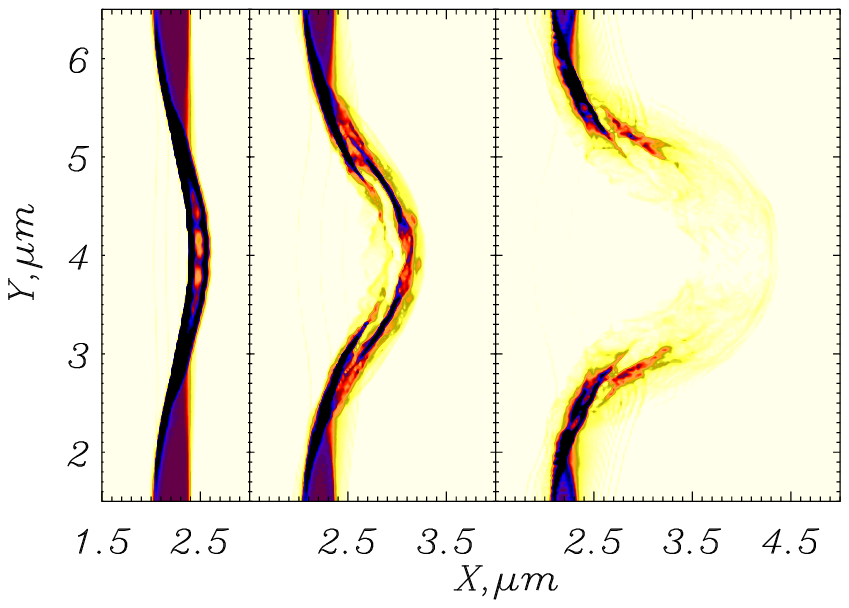}
\includegraphics[width=0.55\textwidth]{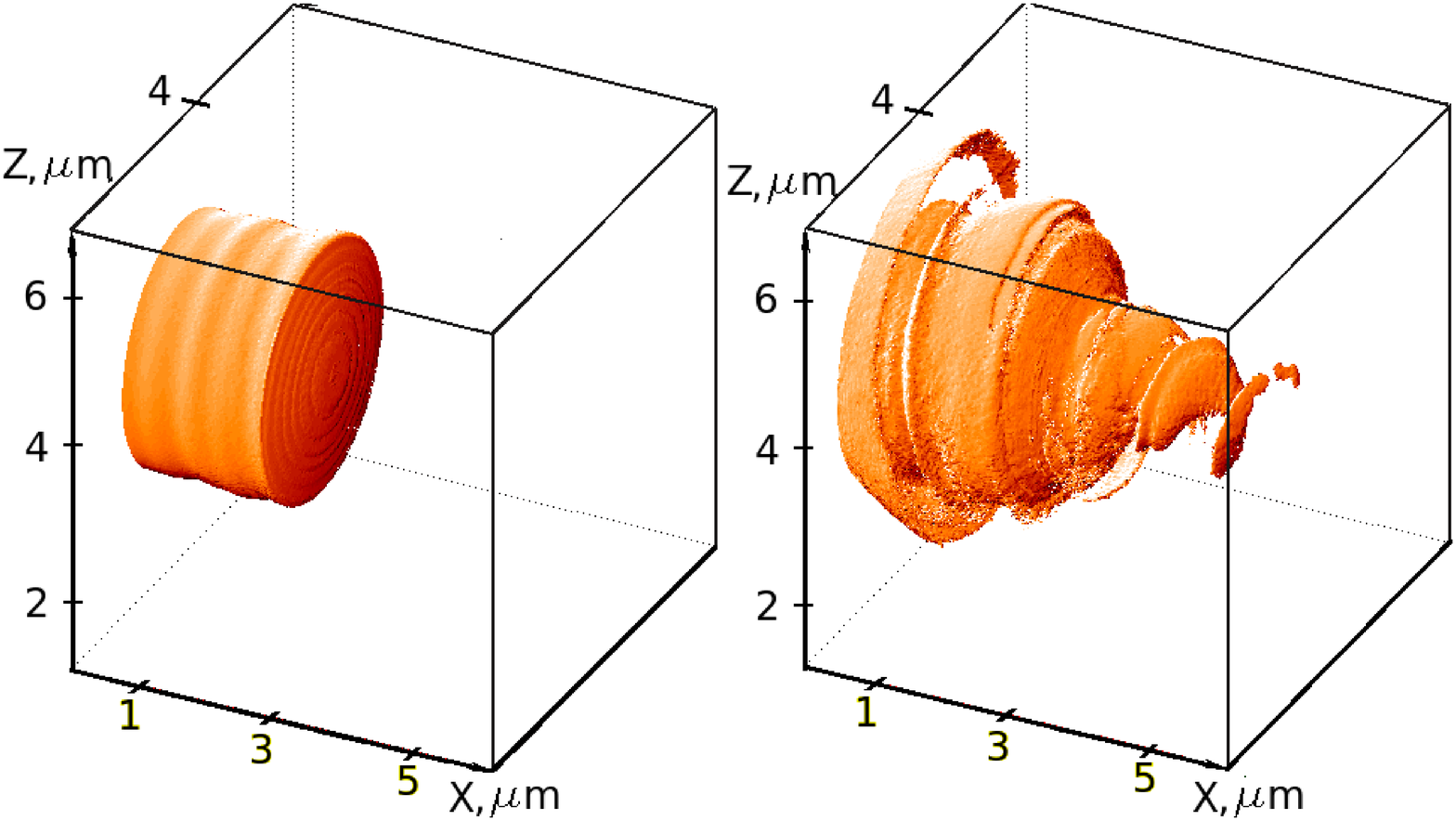}
\caption{The distribution of ion density at $t=75, 100$ and $130$ fs and the electromagnetic energy at $t=50$ fs and $t=100$ fs in the case of tightly focused (3$\mu$m width) laser pulse interaction with an ultrathin foil.
\label{fig:3D2}}
\end{figure*}
In Fig.~\ref{fig:3D2} the results of the interaction of a
tightly focused (3 $\mu$m width) Gaussian laser pulse with a target of density 
$n_e=9 n_{cr}=1\times 10^{24}$ cm$^{-3}$ and thickness of 0.4 $\mu$m are shown.
In this case the pulse focusing was tight enough to contribute dramatically
to the induced transparency of the target.

In both cases presented here the density and the width of the targets were
chosen in a way to ensure their opacity on the basis of the
1D analysis. However, 3D effects decrease the transparency
threshold because the foil tends to expand in perpendicular direction.
For the tightly focused Gaussian laser pulse this effect 
is very pronounced
so that the foil became transparent even if initially it was opaque.
Since the use of a target with densities not very far from the
transparency threshold is more suitable to achieve an efficient
acceleration rate, the shape of the laser pulse becomes a critical issue and
the use of ``flat-top'' laser pulses, whenever possible,
may help.

\ack
This work was supported by CNR-INFM and CINECA (Italy) through
the super-computing initiative and by CNR via a RSTL project.
Some of the simulations were performed on the Linux Cluster of MPI-K, 
Heidelberg. Part of the work was performed during a stay of two of the authors
at Queen's University, Belfast, UK, supported by a Visiting Research
Fellowship (A.M.) and by COST-P14 (S.T.). T.V.L. also
acknowledges support from RFBR (via 08-02-08244 grant).
We are grateful to D. Bauer, F. Cornolti and F. Pegoraro for critical
reading and comments.

\section*{References}

\bibliographystyle{unsrt}
\bibliography{paper_revised_am2}

\end{document}